\documentclass[aps,pre,showpacs,showkeys,superscriptaddress,nofootinbib,groupedaddress,notitlepage]{revtex4-1}  

\usepackage{graphicx}
\usepackage{bm}
\usepackage{enumerate}
\usepackage{mathtools}

\usepackage{color}

\usepackage{pifont}

\begin{document}

\title{An introduction to the derivation of surface balance equations without the excruciating pain }

\author{Alexandre Martin}
\thanks{Associate Professor, Department of Mechanical Engineering}
\email{alexandre.martin@uky.edu}
\affiliation{University of Kentucky, Lexington, KY, 40506}
\author{Huaibao Zhang}
\thanks{Research Associate, Computational Fluid Dynamics Research Center, School of Physics}
\email{zhanghb28@mail.sysu.edu.cn}
\affiliation{Sun Yat-Sen University, Guangzhou, China, 510006}
\author{Kaveh A. Tagavi}
\thanks{Professor, Department of Mechanical Engineering}
\email{kaveh.tagavi@uky.edu}
\affiliation{ University of Kentucky, Lexington, KY, 40506 }

\begin{abstract}
Analyzing complex fluid flow problems that involve multiple coupled domains, each with their respective set of governing equations, is not a trivial undertaking. Even more complicated is the elaborate and tedious task of specifying the interface and boundary conditions between various domains. This paper provides an elegant, straightforward and universal method that considers the nature of those shared boundaries and derives the appropriate conditions at the interface, irrespective of the governing equations being solved. As a first example, a well-known interface condition is derived using this method. For a second example, the set of boundary conditions necessary to solve a baseline aerothermodynamics coupled plain/porous flow problem is derived. Finally, the method is applied to two more flow configurations, one consisting of an impermeable adiabatic wall and the other an ablating surface.

\end{abstract}


\maketitle

\section{Introduction}
Arguably, one of the more complex and painful~\cite{pain_grad} 
tasks in developing a numerical simulation code, such as those in the field of Computational Fluid Dynamics (CFD), is constructing  the boundary condition equations. The task becomes even more complicated when multiple codes that solve different sets of governing equations need to be coupled via physical boundary conditions at their interfaces.

The traditional approach for coupling different codes is to develop the so-called ``balance equations" at the interface, which are then used to solve for various parameters at that particular interface. These balance equations are often determined either by one or a combination of the following methodologies:
\begin{enumerate}[(a)]
\setlength{\itemindent}{10.5pt}
\item determination of empirical relations followed by phenomenological derivation of transport equations; 
\item  thorough consideration of certain physical phenomenon at the interface.
\end{enumerate}

\begin{figure}
\vspace{-0.4in}
\resizebox{3.in}{!}{\includegraphics{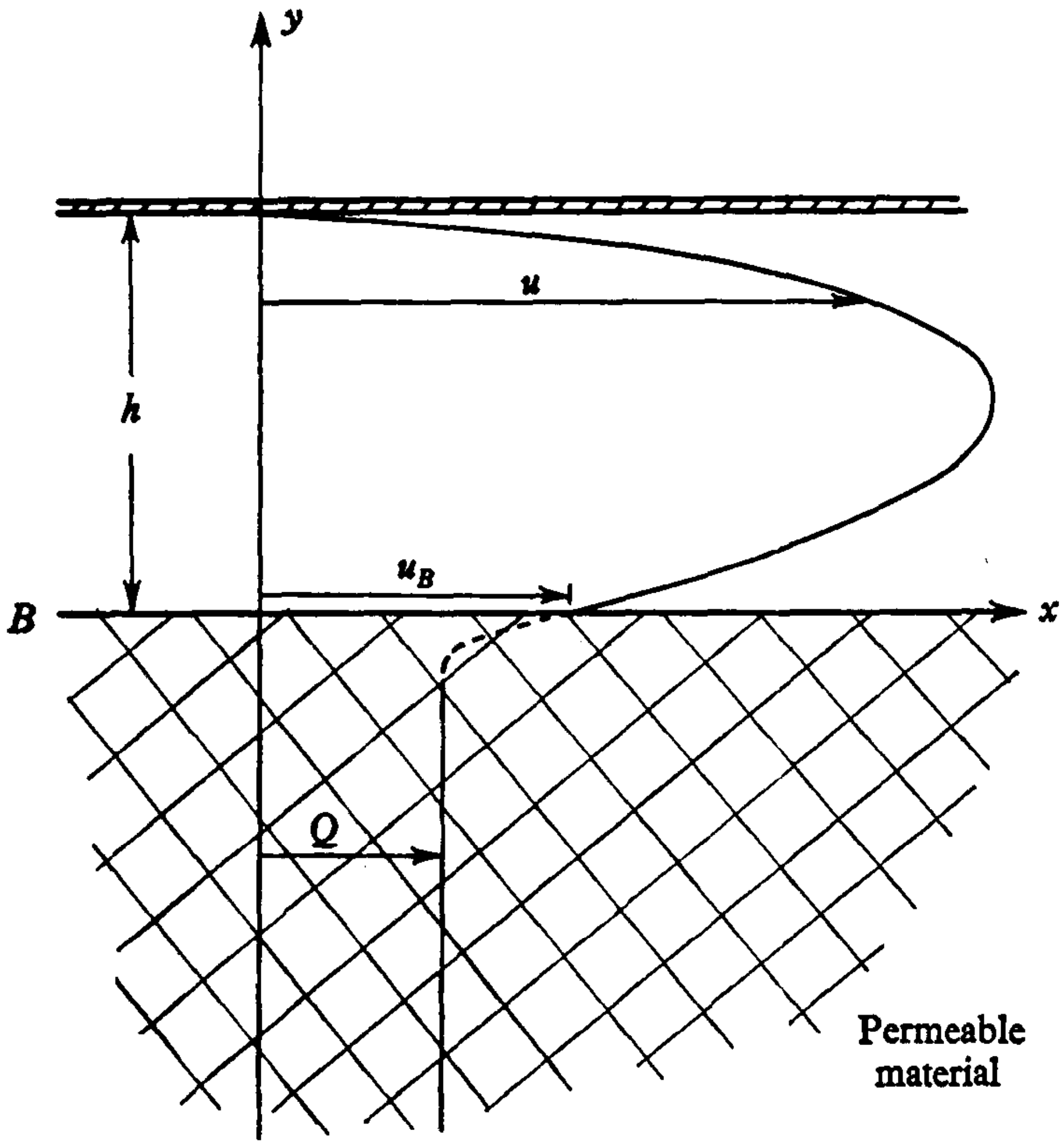}}
\caption{\small {\it Velocity profile of a plain flow over a porous material (image taken from~\citet{Beavers:1967aa} reproduced with permission)}}
\label{beavjo}
\end{figure}

 A common and oft-used example of balance condition of type (a) was developed in the late 1960s by~\citet{Beavers:1967aa}. That boundary condition, obtained experimentally, governs the behavior of the velocity at the interface between a  plain flow parallel to a porous medium,\footnote{
The curious reader is invited to learn about the history of the Beavers-Joseph equation by reading the well-documented paper of~\citet{Nield:2009aa}.} as illustrated in Fig.~\ref{beavjo}. The boundary condition at interface $B$ is:
\begin{equation}
\left.\frac{\partial u}{\partial y}\right|_{B} = \frac{\alpha_{BJ}}{\sqrt{K}} (u_{B} - Q) \ .
\end{equation}
The solution of this partial differential equation determines the ``slip" velocity $u$  at the interface, which is a function of the permeability $K$, the superficial velocity $Q$ in the porous media, and the Beavers-Joseph parameter $\alpha_{BJ}$. This equation was later given a ``statistical treatment" by~\citet{Saffman1971aa}, where it was shown that the velocity $Q$ could be neglected to obtain:
\begin{equation}
\left.\frac{\partial u}{\partial y}\right|_{B} = \frac{\alpha_{BJ}}{\sqrt{K}} u_B \ .
\label{Jones}
\end{equation}
A few years later~\cite{Jones1973aa}, it was demonstrated that this equation could be generalized by relating the velocity at the wall to the shear stress:
\begin{equation}
\left[\frac{\partial u}{\partial y} + \frac{\partial v}{\partial x}\right]_{B}   = \frac{\alpha_{BJ}}{\sqrt{K}} u_B \ .
\label{Saffman}
\end{equation}

A common example of type (b) boundary condition derivation is when a CFD code is coupled with a Material Response (MR) solver for studying aerothermodynamics problems, such as the surface balance equations presented by~\citet{Rasky}.

In this paper a systematic method of writing complex boundary conditions is presented. In order to demonstrate the feasibility of the method, first, a simple but well-established flow is used. The method is then applied to the much more complex problem of an aerothermodynamic flow coupled to a porous flow to demonstrate the versatility of the method. 

\section{Flux-balancing at the boundary}\label{flux_balancing}

As mentioned above, the modeling of surface phenomena between adjoining domains of different flow types has proven to be challenging from the computational point of view. This happens when two independent solvers are specifically designed to solve different sets of equations  where each is applicable only to their respective domain. Coupling two different sets of governing equations at the interface is not a trivial matter. The difficulty comes from two sources: (a) mathematically -- the non-linearity and order of each set of differential equations result in a complex system to solve, and (b) numerically -- the implementation of the resulting interface condition may require advanced cell mapping and reconstruction techniques, especially if the coupling scheme is used in a multi-processor environment.

The previous works on this topic have focused on two main types of applications. The first type consists of incompressible, non-reacting flows~\cite{OchoaTapia1995vn,OchoaTapia1995aa, Cieszko:1999uq,Saffman1971aa, a1:2006aa} which results in a simplified system of equations where the mass and energy equations need not be explicitly solved. For this type of applications, the problem is simplified to determining values for pressure, normal velocity, and both tangential velocities at the interface. In most cases, the condition at the interface is obtained for the very specific flows and cannot be extended to other flow configurations. A well-known example of such condition is the Beavers-Joseph equation, which, for instance, is not applicable to flows where the normal velocity component at the interface cannot be neglected.

The second type of application addresses compressible, reacting flows: for example, the interaction of an atmospheric flow over the porous heat shield of a re-entry vehicle~\cite{martinJTHT2010,martinJTHT2009,kuntz}. Previous works on surface coupling have been based on a methodical analysis of all physical phenomena, which is then presented as surface balance equations~\cite{Rasky}. Again, such an approach has the drawback of not being universal, varying from case to case depending on the equations that are solved in the respective domains. Moreover, certain phenomena are sometimes neglected without proper justification (e.g., the shear stress term in the energy equation~\cite{martinJTHT2009}), or an entire conservation equation is neglected (e.g. the momentum equation reduced to zero pressure gradient~\cite{Rasky}).

However, instead of trying to document every possible pair of physical phenomenon that could occur at an interface, the present paper highlights an elegant and straightforward 
method to model surface equations: let the governing equations do the work. 

Nearly all governing equations are composed of three types of terms: the time dependent storage term, the flux term, and the source term. Both the time-dependant and the source terms are locally evaluated. By contrast, the flux terms are based on the transport of an intensive property through the enclosed boundaries of a volume. This principle 
has been defined by Reynolds, in his classic Reynolds Transport Theorem~\cite{Reynolds1903aa}:
\begin{equation}
 \int_V  \frac{\partial  {\bf Q}}{\partial t}  {dV} + \oint_{A} {{\bf Q}({\bf u} \cdot {\bf n})} {dA}=  \int_V  {\bf S} {dV}
\label{RTT}
\end{equation}
where $\bf u$ is the velocity vector. In this equation, the first term represents the time-dependent storage term, the second, the flux term, and the third, the source term. In layman's term: the change in time of quantity $\bf Q$  in volume $V$ is given by the net amount of the quantity that enters through the boundaries $A$, plus any source terms that may exist within the control volume. 
Equation~\eqref{RTT} can be re-written by defining a flux matrix ${\pmb{ \mathcal F}}$ that includes the product $\bf Q$ and  $\bf u$, as well as other surface phenomena (e.g., pressure contributions in the momentum equation):
\begin{equation}
 \int_V  \frac{\partial  {\bf Q}}{\partial t}  {dV} + \oint_{A} {{\pmb{\mathcal F}} \cdot {\bf n}} {dA}=  \int_V  {\bf S} {dV}
\label{RTT_flux}
\end{equation}

If two domains -- domain \ding{192}  and  domain \ding{193} --  are connected via surface $A_B$, the quantity $\pmb{\mathcal F}$ leaving domain \ding{192} through area $A_B$ is exactly the same as that entering domain \ding{193}. 
This  concept, applied to Eq.~\eqref{RTT_flux}, simply means that independent of the storage and source terms in each respective domain, the flux term on each side of the interface must be equal to each other. 
This principle can be expressed mathematically by:
\begin{equation}
 \left[ \int_{A_B} {{\pmb{ \mathcal F}} \cdot {\bf n}} {dA}  \right]_\text{\ding{192}} =  \left[ \int_{A_B} {{\pmb{ \mathcal F}} \cdot {\bf n}} {dA}\right]_\text{\ding{193}}
 \label{fluxmatching}
\end{equation}
From this analysis, it can be concluded that when imposing a boundary condition at the interface of two domains, Eq.~\eqref{fluxmatching} must be satisfied. This  basic principle is further explained in the following sections, using two classic problems as examples.

It is to be noted that this derivation does not include surface source terms, some of which could take place at the interfaces. In such a case, Eq.~\eqref{fluxmatching} is rewritten as:
\begin{equation}
 \left[ \int_{A_B} {{\pmb{ \mathcal F}} \cdot {\bf n}} {dA}  \right]_\text{\ding{192}} =  \left[ \int_{A_B} {{\pmb{ \mathcal F}} \cdot {\bf n}} {dA}\right]_\text{\ding{193}}+  \int_{A_B}  {\bf S_A} {dA}
 \label{surface_s}
\end{equation}
where $\bf S_A$ is the rate of the intrinsic quantity being added to the control surface $A_B$.

\section{Example 1: coupling of Darcy's law and Stokes equations}
To illustrate the concept of ``flux-balancing",  as represented in Eq.~\eqref{fluxmatching} above, and to verify and validate its feasibility, the method is first applied to a classical case with the physical configuration illustrated in Fig.~\ref{beavjo}, which was used to derive the Beavers-Joseph equation. For this type of configurations, the porous flow is solved using Darcy's law, and  the plain flow using the Stokes equations, as detailed below.

\subsection{Darcy's law}
A porous medium is usually defined as a fixed-in-space solid matrix with connected void spaces through which a fluid can flow~\cite{Nield:2aa}. The law governing the flow thorough a porous medium was obtained experimentally by Henry Darcy in 1856, while  observing the filtration system of the aqueducts of the city of Dijon~\cite{Darcyslaw}. Darcy's law is given by 
\begin{equation}
\frac{\partial p}{\partial x} =-\frac{\mu}{K} v_s,
\label{Eqn:darcy}
\end{equation}
where $\mu$ is the viscosity of the fluid and $K$ is the permeability of the porous medium. The superficial velocity, $v_s$, is calculated by averaging the velocity over a cross section along the porous medium. 

Although the rigorous way to derive Darcy's law is to average the Navier-Stokes equation over the porous volume~\cite{whitaker}, it is also possible 
to derive it using scale analysis. Assuming that the flow is Newtonian, steady-state, one-dimensional, and incompressible, the momentum equation in the $x$-direction, reduces to
\begin{equation}
0 =  -\frac{\partial p}{\partial x}
	  + \frac{\partial}{\partial y }\left[\mu
	      \frac{\partial u}{\partial y}\right] \ . 
\label{eqn:simple_mom}
\end{equation}
The solution of this equation with a no-slip velocity boundary condition is a Poiseuille flow, and takes the form of:
\begin{equation}
{u}=-\dfrac{\partial p}{\partial x} \frac{(hy-y^2)}{2\mu} \ , 
\label{eqn:pipeave1}
\end{equation}
where $h$ is the height (in $y$ direction) of the uniform cross-section of the flow enclosure.    
An expression for the superficial velocity in the $x$-direction can be obtained by integrating over the cross section:
\begin{equation}
 v_s =-\dfrac{\partial p}{\partial x} \frac{h^2}{12\mu} \ .
\label{eqn:pipeave}
\end{equation}
By comparing Eq.~\eqref{Eqn:darcy} and Eq.~\eqref{eqn:pipeave}, it can be seen that a permeability can be derived as $K=h^2/12$. Going back to Eq.~\eqref{eqn:simple_mom}, it can therefore be deduced that the characteristic length of the derivative is proportional to the inverse of the square root of $K$, that is $\frac{\partial }{\partial y}\sim \frac{1}{\sqrt{K}} $ or more specifically $\frac{\partial }{\partial y} = -  \frac{\alpha}{\sqrt{K}}$, where $\alpha$ is a proportionality constant. Utilizing the same flux form as for the plain flow, the 2D Darcy's laws can be written as: 
\begin{equation}
	\nabla \cdot   
	\begin{pmatrix}
		p & 0   \\
		0 & p  \\
	\end{pmatrix}-
	\begin{pmatrix}
		-\frac{\mu}{K} u    \\
		 -\frac{\mu}{K} v \\
	\end{pmatrix}
	=0
\quad. 
\end{equation}
Using the result from  the above scale analysis and setting $ \frac{\partial }{\partial x} =  \frac{\partial }{\partial y}= \frac{- \alpha_s}{\sqrt{ K}}$, this equation can be formulated as:
\begin{equation}
	\nabla \cdot   
	\begin{pmatrix}
		p & 0   \\
		0 & p  \\
	\end{pmatrix}
	-\nabla \cdot \alpha_s \mu
	\begin{pmatrix}
		 2  \frac{u }{\sqrt{K}}  &  \frac{ u + v}{\sqrt{K}}     \\
		 \frac{ v+u}{\sqrt{K}}  & 2  \frac{ v}{\sqrt{K}} \\
	\end{pmatrix}
	=0
\quad. 
\label{eqn:darcy_flux_before}
\end{equation}
After integrating over volume $V$, and applying the divergence theorem, the following is obtained:
\begin{equation}
    \oint \left[
	\begin{pmatrix}
		p & 0   \\
		0 & p  \\
	\end{pmatrix}
	\cdot {\bf n}- \alpha_s \mu
	\begin{pmatrix}
		 2 \frac{u}{\sqrt{K}}  &  \frac{ u + v}{\sqrt{K}}     \\
		  \frac{ v+u}{\sqrt{K}} & 2  \frac{ v}{\sqrt{K}}  \\
	\end{pmatrix}
	\cdot {\bf n} \right]  dA =0
\quad. 
\label{eqn:darcy_flux}
\end{equation}
where $\bf n$ is the normal vector of surface $A$ bounding volume $V$. In this formulation, Eq.~\eqref{eqn:darcy_flux} therefore describes momentum fluxes that leaves (or enters) a porous volume $V$ through surface $A$.

\subsection{Stokes equations}
For the proposed example, the plain flow is governed by the Stokes equations in 2D, assuming a Newtonian fluid. In matrix form, the steady state form of these equations becomes:
 \begin{equation}
	\nabla \cdot   
	\begin{pmatrix}
		p & 0   \\
		0 & p  \\
	\end{pmatrix}
	-\mu \nabla \cdot   
	\begin{pmatrix}
		2 \frac{\partial u}{\partial x} &  \frac{\partial u}{\partial y}+\frac{\partial v}{\partial x}   \\
		 \frac{\partial v}{\partial x} +  \frac{\partial u}{\partial y} & 2\frac{\partial v}{\partial y}  \\
	\end{pmatrix}
	=0
	\quad. 
\end{equation}
Similarly, this equation can be expressed as a momentum flux by integrating over volume $V$, while applying the divergence theorem:
\begin{equation}
\oint   \left[
	\begin{pmatrix}
		p & 0   \\
		0 & p  \\
	\end{pmatrix}
	\cdot {\bf n}-\mu 
	\begin{pmatrix}
		2\frac{\partial u}{\partial x} &  \frac{\partial u}{\partial y}+ \frac{\partial v}{\partial x}   \\
		 \frac{\partial v}{\partial x} +  \frac{\partial u}{\partial y} & 2 \frac{\partial v}{\partial y}  \\
	\end{pmatrix}
	\cdot {\bf n}\right]dA=0
\quad,
\label{eqn:stokes_flux}
\end{equation}

\subsection{Flux-balancing: parallel flow over a permeable wall}
It can be shown that the Beavers-Joseph boundary condition can be derived analytically by coupling the plain flow and porous flow at the interface. For the interface between the plain flow and the porous medium, the fluxes on both sides must be balanced, i.e. no fluxes are destroyed or created at the surface and that the fluxes only pass through the interface. Because the two adjacent domains share a contact interface of area $A_B$ and a normal vector $\bf n_B$, an equation at the interface is obtained by ``balancing", or equating, the fluxes from Eqs.~\eqref{eqn:stokes_flux} and \eqref{eqn:darcy_flux}:
\begin{equation}
 \left[
	\begin{pmatrix}
		p & 0   \\
		0 & p  \\
	\end{pmatrix}-
	\begin{pmatrix}
		2\mu \frac{\partial u}{\partial x} & \mu \frac{\partial u}{\partial y}+\mu \frac{\partial v}{\partial x}   \\
		\mu \frac{\partial u}{\partial y}+\mu \frac{\partial v}{\partial x} & 2\mu \frac{\partial v}{\partial y}  \\
	\end{pmatrix}
	  \right]_f  
	\cdot {\bf n_B} =
	\left[
	\begin{pmatrix}
		p & 0   \\
		0 & p  \\
	\end{pmatrix} -
	\begin{pmatrix}
		 2 \alpha_s \frac{ \mu}{\sqrt{K}} u & \alpha_s \frac{ \mu}{\sqrt{K}} (u + v)    \\
		 \alpha_s \frac{ \mu}{\sqrt{K}} (v+u) & 2 \alpha_s \frac{ \mu}{\sqrt{K}} v \\
	\end{pmatrix}
	\right]_p
	\cdot {\bf n_B}
\label{eqn:BJ-flux-equal}
\end{equation}
Since the plain flow is overlaid on top of the porous medium, as illustrated in Fig.~\ref{beavjo}, then ${\bf n_B}=(0,-1)$, and the coupling equation becomes:
\begin{subequations}
\begin{align}
\label{subeqn:20a}
\left[ \frac{\partial u}{\partial y}+ \frac{\partial v}{\partial x} \right]_f &= \alpha_s \frac{1 }{\sqrt{K}} (u+v)\Bigg|_p\\
\left[-p+2\mu \frac{\partial v}{\partial y}\right]_f  &= \left[-p +2 \alpha_s \frac{ \mu}{\sqrt{K}} v \right]_p
\end{align}
\end{subequations}
where the subscript $f$ and $p$ denote the fluid and porous medium sides, respectively. It can easily be seen that Eq.~\eqref{subeqn:20a} is the same as Eq.~\eqref{Saffman}. If it is assumed that the velocity in the $y$-direction is zero,  the above set of equations reduces to:
\begin{subequations}
\begin{align}
\frac{\partial u_f}{\partial y} &=  \frac{\alpha_s  }{\sqrt{K}} u_p\label{bj_dera}\\
p_f&=p_p\label{bj_derb}
\end{align}
\label{bj_der}
\end{subequations}

\noindent Equation~\eqref{bj_dera} is exactly the Beavers-Joseph boundary condition as expressed in Eq.~\eqref{Jones}. The present method also yields the continuity of pressure at the interface, Eq.~\eqref{bj_derb}, which is implicitly assumed in most boundary condition treatments. This derivation, however, clearly shows that this equation is only valid for very specific types of flow (Stokes, in the plain flow domain, Darcy, in the porous media domain), for a very specific geometrical configuration (stratified flow over a porous medium), with a very specific assumption (no velocity in the $y$-direction). It is clear that this interface condition would not be appropriate to couple, for example, the Navier-Stokes equation with the Forchheimer equation for porous flow~\cite{Forchheimer1901aa} since it is missing some of the key terms, such as the advection term.

\section{Example 2: Coupling of Darcy-Brinkman and Navier-Stokes equations}
Another common flow situation is that of a porous material interacting with a reacting compressible flow. One typical example of such situation is when a spacecraft enters an atmosphere while traveling at hypersonic speeds~\cite{martinJTHT2009-1,martinJOPCS2011,martinJTHT2009}. Under such extreme conditions, the spacecraft'€™s thermal protection system (TPS) experiences high levels of aerodynamic heat from the flow around it. One type of TPS that is widely used is a charring ablator TPS, a part of which ablates at high temperatures. As a result, relatively cooler pyrolysis gases are introduced into the flow field while leaving behind a carbon matrix forming a porous medium. Traditionally, a CFD code and a Material Response (MR) code are used to model the flow field and the TPS side, respectively. To fully and accurately analyze an atmospheric entry, however, a fully-coupled approach that can allow complete solutions for both the material and the fluid is desired. Such full scale analysis has not been performed yet. Both set of governing equations in the CFD and MR solvers can be written in the following form
\begin{equation}
\frac{\partial \bf Q}{\partial t} + {\bf \nabla} \cdot (\pmb{\mathcal F}_\text{adv} -\pmb{\mathcal F}_\text{diff}) =  {\bf S}
\label{eq:gov1},
\end{equation}
where $\bf Q$ is a vector of the conserved variables, ${\pmb{\mathcal F}_\text{adv} }$ and $\pmb{\mathcal F}_\text{diff}$ are advective and diffusive fluxes, respectively, and $\bf S$ is a source vector. Integrating this last equation over a control volume results into Eq.~\eqref{RTT}. 

\subsection{Fluid Dynamics Governing Equations}
For a reacting compressible flow, the terms of Eq.~\eqref{eq:gov1} are:
\begin{align}
	{\bf Q} = 
	\begin{pmatrix}
		\rho_{g_1} \\
		\vdots\\
		\rho_{g_\text{ngs}} \\
		\rho_g u \\
		\rho_g v \\
		\rho_g w \\
		\rho_g e
	\end{pmatrix}
	\;
	\pmb{\mathcal F}_\text{adv} = 
	\begin{pmatrix}
		\rho_{g_1} u & \rho_{g_1} v & \rho_{g_1} w   \\
		\vdots & \vdots & \vdots \\
		\rho_{g_\text{ngs}} u & \rho_{g_\text{ngs}} v & \rho_{g_\text{ngs}} w   \\
		\rho_g u^2 + p  &  \rho_g v u & \rho_g w u \\
		\rho_g u v & \rho_g v^2 + p & \rho_g w v \\
		\rho_g u w &\rho_g v w   & \rho_g w^2 + p \\
		\rho_g u h & \rho_g v h & \rho_g w h
	\end{pmatrix}
	\;
	\pmb{\mathcal F}_\text{diff} = 
	\begin{pmatrix}
		\multicolumn{1}{c}{ -{\bf J}_1 }  \\
		\multicolumn{1}{c}{ \vdots }  \\
		\multicolumn{1}{c}{ -{\bf J}_\text{ngs} }  \\
		\tau_{xx} \qquad \tau_{xy} \qquad \tau_{xz} \\
		\tau_{yx} \qquad \tau_{yy} \qquad \tau_{yz} \\
		\tau_{zx} \qquad \tau_{zy} \qquad \tau_{zz} \\
		\multicolumn{1}{c}{ {\pmb \tau} {\bf u}  - {\bf q}  -\sum_{i=1}^\text{ngs} ({\bf J}_i h_i) }
	\end{pmatrix}
	\; 
	{\bf S} = 
	\begin{pmatrix}
		\dot{\omega}_1 \\
		\vdots\\
		\dot{\omega}_\text{ngs} \\
		0 \\
		0 \\
		0 \\
		0
	\end{pmatrix}
	\label{fluid_flux}
\end{align}
where $ \pmb \tau$ is the viscous tensor, ${\bf J}_1,\ldots,{\bf J}_\text{ngs}$ are mass diffusion for each species, and ${\bf q}$ is the heat conduction vector. The subscript $i$, from 1 to $ngs$ (total number of gas species), represents each gas species. In these expressions, $\rho_g$ is overall gas density, ${\bf u} = (u, v, w)$  is the bulk velocity, and $e$ and $h$ are respectively the total energy and total enthalpy of the gas per unit mass.

\subsection{Material Response Governing Equations }
Material Response codes solve for gaseous mass, solid mass, momentum, and energy conservation equations.
Thus, the terms in Eq.~\eqref{eq:gov1} for a porous medium are given by:

\begin{align}
	{\bf Q} = 
	\begin{pmatrix}
		\phi\rho_{g_1} \\
		\vdots\\
		\phi\rho_{g_\text{ngs}} \\
		\rho_{s_1} \\
		\vdots\\
		\rho_{s_\text{ngs}} \\		
		\phi\rho_g u \\
		\phi\rho_g v \\
		\phi\rho_g w \\
		\phi\rho_g e + \rho_s e_s
	\end{pmatrix}
	\;
\pmb{\mathcal F}_\text{adv} = 
	\begin{pmatrix}
		\phi \rho_{g_1} u & \phi \rho_{g_1} v & \phi \rho_{g_1} w\\
		\vdots & \vdots & \vdots \\
		\phi \rho_{g_\text{ngs}} u & \phi \rho_{g_\text{ngs}} v & \phi \rho_{g_\text{ngs}} w\\
		0 & 0 & 0 \\
		\vdots & \vdots & \vdots \\
		0 & 0 & 0 \\
		\phi \rho_g u^2 + p  	& \phi \rho_g v u 		& \phi \rho_g w u \\
		\phi \rho_g u v 		& \phi \rho_g v^2 + p 	& \phi \rho_g w v \\
		\phi \rho_g u w 		& \phi \rho_g v w  	& \phi \rho_g w^2 + p \\
		\phi \rho_g u h  		& \phi \rho_g v h& \phi \rho_g w h
	\end{pmatrix}
\; 
\pmb{ \mathcal F}_\text{diff} = 
	\begin{pmatrix}
		-{\bf J}_1   \\
		 \vdots   \\
		 -{\bf J}_\text{ngs}   \\
		   \\
		 {\bf 0}   \\
		   \\
		   \\
		 {\bf 0}   \\
		   \\
		- \dot{\bf q}'' -\sum_{i=1}^\text{ngs} ({\bf J}_i h_i)  
	\end{pmatrix}
\;
	{\bf S} = 
	\begin{pmatrix}
		\dot{\omega}_{g_1} \\
		\vdots\\
		\dot{\omega}_{g_\text{ngs}} \\
		\dot{\omega}_{s_1} \\
		\vdots\\
		\dot{\omega}_{s_{nss}} \\
		D_x\\
		D_y\\
		D_z\\
		S_\text{D}
	\end{pmatrix}
	\label{eq:solid_matrices}
\end{align}

where $e_s$ and $\rho_s$ is the total energy and total density in the solid phase, respectively. In the Darcy-Brinkman formulation, the porous medium viscous forces are treated as a source term in the momentum equations,  denoted by $D_x$, $D_y$, and $D_z$ in Eq.~\eqref{eq:solid_matrices}, and may be calculated by solving the following linear equation:
\begin{align}
\begin{pmatrix}
K_{xx} & K_{xy} & K_{xz} \\
K_{yx} & K_{yy} & K_{yz} \\
K_{zx} & K_{zy} & K_{zz} \\
\end{pmatrix}
\begin{pmatrix}
D_{x}  \\
D_{y}  \\
D_{z}  \\
\end{pmatrix}
=-\phi\mu
\begin{pmatrix}
u  \\
v  \\
w  \\
\end{pmatrix},
\label{eq:Darcy source}
\end{align}
The three-by-three matrix on the left hand side is an anisotropic tensor of solid permeability.
If the material is orthotropic, Eq.~\eqref{eq:Darcy source} can be greatly simplified:
\begin{align}
	\begin{pmatrix}
	K_{xx} & 0 &0  \\
	0 & K_{yy} &0  \\
	 0&0  &K_{zz} \\
	\end{pmatrix}
	\begin{pmatrix}
	D_{x}  \\	D_{y}  \\	D_{z}  \\
	\end{pmatrix}
	=-\phi\mu
	\begin{pmatrix}
	u  \\	v  \\	w  \\
	\end{pmatrix},
\end{align}
\begin{align}
	D_x = -\frac{\phi\mu}{K_{xx}} u, \qquad
	D_y = -\frac{\phi\mu}{K_{yy}} v, \qquad
	D_z = -\frac{\phi\mu}{K_{zz}} w.
	\label{eq:simplified 1}
\end{align}
The diffusive effect of porous media is modeled as a source term in the energy equation, which is given as:
\begin{equation}
    S_\text{D}  = D_x u + D_y v + D_z w
\end{equation}



As was done in the previous example with Eq.~\eqref{eqn:darcy_flux_before}, a dimensional analysis moves back the Darcy forces and work from the source term to the diffusive fluxes:
\begin{align}
\pmb{\mathcal F}_\text{adv} = 
	\begin{pmatrix}
		\phi \rho_{g_1} u & \phi \rho_{g_1} v & \phi \rho_{g_1} w\\
		\vdots & \vdots & \vdots \\
		\phi \rho_{g_\text{ngs}} u & \phi \rho_{g_\text{ngs}} v & \phi \rho_{g_\text{ngs}} w\\
		0 & 0 & 0 \\
		\vdots & \vdots & \vdots \\
		0 & 0 & 0 \\
		\phi \rho_g u^2 + p  	& \phi \rho_g v u 		& \phi \rho_g w u \\
		\phi \rho_g u v 		& \phi \rho_g v^2 + p 	& \phi \rho_g w v \\
		\phi \rho_g u w 		& \phi \rho_g v w  	& \phi \rho_g w^2 + p \\
		\phi \rho_g u h 		& \phi \rho_g v h		& \phi \rho_g w h
	\end{pmatrix}
\; 
\pmb{\mathcal F}_\text{diff} = 
	\begin{pmatrix}
		& -{\bf J}_1 &  \\
		& \vdots &  \\
		& -{\bf J}_\text{ngs} &  \\
		0&  0 &0  \\
		\vdots & \vdots & \vdots \\
		0&  0 &0  \\
		\\
		& \pmb{\mathcal D}_\text{F} &\\
		\\
				&{\bf W}_\text{F} - \dot{\bf q}'' -\sum_{i=1}^\text{ngs} ({\bf J}_i h_i) &  
	\end{pmatrix}
	\label{eq:solid_matrices2}
\end{align}
where
\begin{align}
\pmb{\mathcal D}_\text{F} = \alpha_s \phi\mu
\begin{pmatrix} 
2\frac{u}{\sqrt {K_{xx}}} & \frac{u}{\sqrt {K_{yy}}}+\frac{v}{\sqrt {K_{xx}}}&\frac{u}{\sqrt {K_{zz}}}+\frac{w}{\sqrt {K_{xx}}} \\
		\frac{v}{\sqrt{K_{xx}}}+\frac{u}{\sqrt {K_{yy}}} & 2\frac{v}{\sqrt{K_{yy}}} &   \frac{v}{\sqrt {K_{zz}}}+\frac{w}{\sqrt {K_{yy}}} \\
		\frac{w}{\sqrt{K_{xx}}}+\frac{u}{\sqrt {K_{zz}}}&  \frac{w}{\sqrt {K_{yy}}}+\frac{v}{\sqrt {K_{zz}}}& 2\frac{w}{\sqrt{K_{zz}}} \\
	\end{pmatrix}
\end{align}
and  where ${\bf W}_\text{F} = \pmb{\mathcal D}_\text{F}  {\bf u}$.
%

\subsection{Flux-balancing: aerothermodynamic flow over a porous material}
Using the technique demonstrated in Section~\ref{flux_balancing}, the balance of fluxes yields:
\begin{equation}
\left[\pmb{\mathcal F}_\text{adv} 
	\cdot {\bf n}-
\pmb{\mathcal F}_\text{diff}
	\cdot {\bf n}  \right]_f  =
\left[\pmb{\mathcal F}_\text{adv} 
	\cdot {\bf n}-
\pmb{\mathcal F}_\text{diff}
	\cdot {\bf n} \right]_p
	\label{balanceII}
\end{equation}
Without making any assumptions on the direction of the flow, the interface between the porous flow and the plain flow can be  arbitrarily oriented in the $x-$direction. The normal vector of that surface is therefore ${\bf n}=(1,0,0)$.

\subsubsection{Mass balance}
Applying Eq.~\eqref{balanceII} to the conservation of mass for species $i$, using the terms of Eqs.~\eqref{fluid_flux} and \eqref{eq:solid_matrices2}, the interface condition is therefore:
\begin{equation}
\left[\rho_{g_i} u +J_{x,i}  \right]_f = \left[\phi \rho_{g_i} u + J_{x,i} \right]_p
\label{consofmass}
\end{equation} 
The overall mass flow rate, obtained by summing over all species, is conserved at the interface, that is:
\begin{equation}
\dot{m}''= \left[\rho_{g} u  \right]_f = \left[\phi \rho_{g} u  \right]_p 
\label{speciessum}
\end{equation}
and the mass diffusion term for the species on the fluid side is calculated by Fick's law as
\begin{equation}
J_{x,i}=- \rho_g D \frac{\partial Y_i}{\partial x}
\label{ficks}
\end{equation}
Combining Eqs.~\eqref{consofmass} with \eqref{ficks} yields the Surface Mass Balance (SMB) equation:
\begin{equation}
 \boxed{\left[ \dot{m}''Y_{i} - \rho_g D \frac{\partial Y_i}{\partial x} \right]_f = \left[ \dot{m}''Y_{i} - \rho_g D \frac{\partial Y_i}{\partial x} \right]_p}
\end{equation}
which does not make any assumptions about the individual contributions of mass transport phenomenon. 
But if the mass diffusion term is neglected on the porous side, Eq.~(5a) of~\citet{Rasky} is obtained:
\begin{equation}
\left[ \rho_g D \frac{\partial Y_i}{\partial x} \right]_f =  \dot{m''}(Y_{i,f} -Y_{i,p}) \ .
\end{equation}
However, it might not always be proper to neglect mass diffusion in the porous medium flow~\cite{martinAIAA2010-1,lachaud2011}. 
It is to be noted that Eq.~(5b) of~\citet{Rasky} can  be obtained by including a surface source term for surface chemical reactions.

\subsubsection{Momentum balance}
Most Material Response (MR) codes do not solve the momentum equation explicitly, but rather implicitly include Darcy's law in energy and mass conservation. Thus, a Surface Momentum Balance (SMoB) equation is usually not necessary to close the system of equations. However, when coupling to a CFD code, a momentum boundary condition needs to be applied. Using the approach adopted in this work, the momentum fluxes at the interface are therefore:
\begin{subequations}
\begin{align}
\Aboxed{\left[ \rho_g u^2 + p  -\tau_{xx} \right]_f  &= \left[  \phi \rho_g u^2 + p - \alpha_s \phi\mu 2\frac{u}{\sqrt {K_{xx}}} \right]_p}\\
\Aboxed{\left[  \rho_g u v  -\tau_{yx} \right]_f  &= \left[  \phi \rho_g v u  - \alpha_s \phi\mu\left(\frac{v}{\sqrt {K_{xx}}} +  \frac{u}{\sqrt {K_{yy}}}\right)  \right]_p}\\
\Aboxed{\left[ \rho_g u w  -\tau_{zx} \right]_f  &= \left[  \phi \rho_g w u  - \alpha_s \phi\mu\left(\frac{w}{\sqrt {K_{xx}}} +  \frac{u}{\sqrt {K_{zz}}}\right) \right]_p}
\end{align}
\label{eqn:full-momentum}
\end{subequations}

This set of equation is much more detailed than the continuity of pressure condition~\cite{suzuki2008,hisashi_2013,ojas2014} that is used in most coupling approaches:
\begin{equation}
 p_f = p_p
\end{equation}
or the slightly more accurate total pressure condition~\cite{gnoffo2009,martinJTHT2010}:
\begin{equation}
\left[ \rho_g u^2 + p\right]_f  =  \left[  \phi \rho_g u^2 + p \right]_p\\
\end{equation}


\subsubsection{Energy balance}
Applying the same technique to the energy terms yields the Surface Energy Balance (SEB) equation:
\begin{equation}
\boxed{
\begin{aligned}
\left[ \rho_g  u h   - (\tau_{xx} u + \tau_{xy} v +\tau_{xz} w) + \dot{q}''_{x} + \sum J_{x,i} h_i \right]_f = \hspace{5cm} \\
\left[ \phi \rho_g u h -\alpha_s\phi\frac{\mu}{\sqrt{K_{xx}}}\left({u^2}+{v^2}+{w^2}\right) +  \dot{q}''_{x}   + \sum J_{x,i} h_i \right]_p
\end{aligned}
}
\end{equation}
In this equation, it is assumed that $ \frac{u}{\sqrt{K_{xx}}} + \frac{v}{\sqrt{K_{yy}}} + \frac{w}{\sqrt{K_{zz}}} \approx 0$, per mass conservation of an incompressible fluid.  As with the SMB, simplifying using the total conservation of mass (i.e., Eq.~\eqref{speciessum}) and neglecting the work performed by shear on both sides of the surface reduces this equation to Eq.~(6) of~\citet{Rasky}:
\begin{equation}
\left[ -k_T \frac{\partial T_T}{\partial x} -  k_v \frac{\partial T_v}{\partial x} -
\sum_i  \rho_g h_i D \frac{\partial Y_i}{\partial x} \right]_f =  \dot{q}''_{x} + \dot{m}''(h_{p} -h_{f})
\end{equation}

It is to be noted that~\citet{Rasky} include radiative fluxes and emission, both omitted in the present derivation. 

\section{Example 3: Navier-Stokes equations with an impermeable, adiabatic wall}
The procedure outlined in Section~\ref{flux_balancing} can also be used to define an adiabatic, impermeable wall boundary condition for the Navier-Stokes equations. For the sake of simplicity, a single species incompressible flow is considered. Starting from Eq.~\eqref{surface_s}, domain  \ding{192} fluxes  are thus represented by the Navier-Stokes equations, and  domain  \ding{193} fluxes are null. For this example, the surface source term $\bf S_A$ is needed for the momentum equations. This term, defined in the following as  $\bf R''$, represent the reaction forces per unit area on the surface (for instance, in aerodynamic applications, the drag, lift, and lateral forces).

The surface conditions are obtained by applying the fluxes of Eq.~\eqref{fluid_flux} to a surface  oriented in the $x$-direction. Therefore, for the mass balance results in:
\begin{equation}
\boxed{\left[\rho_{g} u  \right]_f = 0}
\label{consofmass3}
\end{equation} 
Considering that the gas density cannot be 0, this leads to a zero velocity in the $x$-direction at the surface. The momentum flux balance then becomes:

\begin{subequations}
\begin{align}
\left[ \rho_g u^2 + p  -\tau_{xx} \right]_f  &=  R''_x\\
\left[  \rho_g u v  -\tau_{yx} \right]_f  &= R''_y\\
\left[ \rho_g u w  -\tau_{zx} \right]_f  &= R''_z 
\end{align}
\label{eqn:full-momentum3}
\end{subequations}
and the energy flux becomes:
\begin{equation}
\left[ \rho_g  u h   - (\tau_{xx} u + \tau_{xy} v +\tau_{xz} w) + \dot{q}''_{x}  \right]_f = 0 
\end{equation}

This set of equations represents the condition at the wall, for a Navier-Stokes solution. It is to be noted that no assumptions has been made about the velocity at the wall, which makes these equation valid whether or not	 a no-slip condition is enforced. By inserting the mass conservation equation ($\rho_g u = 0$) into both the momentum and energy equations above, and by enforcing a no-slip velocity (${\bf u} = 0$), a more familiar set of equations is obtained

\begin{subequations}
\begin{align}
\Aboxed{\left[  p  -\tau_{xx} \right]_f  &=  R''_x}\\
\Aboxed{\left[   -\tau_{yx} \right]_f  &= R''_y}\\
\Aboxed{\left[  -\tau_{zx} \right]_f  &= R''_z }
\end{align}
\label{eqn:full-momentum3final}
\end{subequations}

\begin{equation}
\boxed{
\left[ \dot{q}''_{x}  \right]_f \equiv \left[  - k \frac{\partial T}{\partial x} \right]_f  =  0
}
\end{equation}

\section{Example 4: Coupling of Navier-Stokes equations and a surface ablating material}

In the field of aerothermodynamics, it is often necessary to account for the interaction of an ablating surface with a high enthalpy flow. This is especially true when studying thermal protection systems, whether it is from the fluid perspective \cite{Trevino2015aa,Mortensen2016} or from the material response perspective \cite{martinJTHT2009,martinJTHT2010}. For this type of problems, the surface of the material is recessing, and mass, energy and momentum is transferred into the flow field. Therefore, domain \ding{192} is a fluid flow, and domain \ding{193}, a recessing solid wall.

The fact that the surface is recessing introduces an additional flux which was not included in Example 3. This flux has taken various names in the literature, but is more appropriately  described as being a {\em grid advection flux}. In terms of control volume analysis, this flux is obtained by implicitly solving the {\em Geometric Conservation Law}\cite{Thomas1979}, and is often used in CFD codes in order to modify a computational grid while calculating the solution. Because it relates the Lagrangian frame to the Eulerian frame, it is often described as an {\em Arbitrary Lagrangian Eulerian} approach, or ALE. More detail of this technique is given in Ref.\cite{trep2}, and applications to control volume problems are presented in Ref.~\cite{zhang_gcl}. From a physical point of view, the grid movement only affects conserved variables, and are therefore described similarly to the advective fluxes. More details on the procedure necessary to obtain this result can be found in Ref.~\cite{martinJTHT2009}.

Using the nomenclature previously defined, and defining ${\bf u}_g$ as the wall (or grid) velocity vector, the grid convection flux $\pmb{\mathcal F}_\text{grid} $ can be cast as:
\begin{equation}
\pmb{\mathcal F}_\text{grid} = {\bf  Qu}_g
\end{equation}
The fluxes from the fluid domain are therefore composed of the grid advection flux, and both advective and diffusive fluxes of Eq.~\ref{fluid_flux}:
\begin{align}
	\pmb{\mathcal F}_\text{grid}  = 
	\begin{pmatrix}
		\rho_{g_1} u_g & \rho_{g_1} v_g & \rho_{g_1} w_g   \\
		\vdots & \vdots & \vdots \\
		\rho_{g_\text{ngs}} u_g & \rho_{g_\text{ngs}} v_g & \rho_{g_\text{ngs}} w_g   \\
		\rho_g u u_g  &  \rho_g v u_g & \rho_g w u_g \\
		\rho_g u v_g & \rho_g v v_g & \rho_g w v_g \\
		\rho_g u w_g &\rho_g v w_g   & \rho_g w w_g \\
		\rho_g e u_g & \rho_g e v_g & \rho_g e w_g
	\end{pmatrix}
		\;
	\pmb{\mathcal F}_\text{adv} = 
	\begin{pmatrix}
		\rho_{g_1} u & \rho_{g_1} v & \rho_{g_1} w   \\
		\vdots & \vdots & \vdots \\
		\rho_{g_\text{ngs}} u & \rho_{g_\text{ngs}} v & \rho_{g_\text{ngs}} w   \\
		\rho_g u^2 + p  &  \rho_g u v & \rho_g u w  \\
		\rho_g v u & \rho_g v^2 + p & \rho_g v w \\
		\rho_g w u &\rho_g  w v   & \rho_g w^2 + p \\
		\rho_g h u & \rho_g h v & \rho_g h w
	\end{pmatrix}
	\;
	\pmb{\mathcal F}_\text{diff} = 
	\begin{pmatrix}
		\multicolumn{1}{c}{ -{\bf J}_1 }  \\
		\multicolumn{1}{c}{ \vdots }  \\
		\multicolumn{1}{c}{ -{\bf J}_\text{ngs} }  \\
		\tau_{xx} \qquad \tau_{xy} \qquad \tau_{xz} \\
		\tau_{yx} \qquad \tau_{yy} \qquad \tau_{yz} \\
		\tau_{zx} \qquad \tau_{zy} \qquad \tau_{zz} \\
		\multicolumn{1}{c}{ {\pmb \tau} {\bf u}  - \dot{\bf q}''  -\sum_{i=1}^\text{ngs} ({\bf J}_i h_i) }
	\end{pmatrix}
	\label{fluid_flux_mofving}
\end{align}
From the material side, the surface fluxes can simply be obtained by setting the gas velocity to 0 in Eq.~\ref{eq:solid_matrices}:
\begin{align}
	\pmb{\mathcal F}_\text{grid}  = 
	\begin{pmatrix}
		{\rho_s}_1 u_g &  {\rho_s}_1 v_g &   {\rho_s}_1 w_g  \\
		&\vdots\\
		{\rho_s}_\text{ngs} u_g &  {\rho_s}_\text{ngs} v_g &   {\rho_s}_\text{ngs} w_g  \\
		 &{\bf 0}  \\
		 &{\bf 0}  \\
		 &{\bf 0}  \\
		\rho_s e_s u_g &  \rho_s e_s v_g &   \rho_s e_s w_g
	\end{pmatrix}
	\;
\pmb{\mathcal F}_\text{adv} = 
	\begin{pmatrix}
		 {\bf 0}  \\
		 \vdots   \\
		 {\bf 0} \\
		 {\bf 0}   \\
		 {\bf 0}   \\
		 {\bf 0}   \\
		 {\bf 0}   \\
	\end{pmatrix}
\; 
\pmb{ \mathcal F}_\text{diff} = 
	\begin{pmatrix}
		 {\bf 0}  \\
		 \vdots   \\
		 {\bf 0} \\
		 {\bf 0}   \\
		 {\bf 0}   \\
		 {\bf 0}   \\
		 {-\dot{\bf q}''}
	\end{pmatrix}
	\label{mat_flux_moving}
\end{align}
These fluxes can thus be used in Eq.~\eqref{surface_s} to obtain the condition at the surface. Similar to the previous examples, a surface oriented in the $x$-direction, and a surface reaction force $\bf R''$ are used. For the mass balance, the following surface condition is obtained:
\begin{equation}
\boxed{
\left[\rho_{g_i} (u-u_g) +J_{x,i}  \right]_f = \left[- \rho_{i} u_g  \right]_s
}
\label{consofmass4}
\end{equation} 
For the momentum balance, the equation becomes:
\begin{subequations}
\begin{align}
\Aboxed{\left[ \rho_g u (u-u_g) + p  -\tau_{xx} \right]_f  &=  R''_x}\\
\Aboxed{\left[  \rho_g v  (u-u_g) -\tau_{yx} \right]_f  &= R''_y}\\
\Aboxed{\left[ \rho_g w (u-u_g)  -\tau_{zx} \right]_f  &= R''_z}
\end{align}
\label{eqn:full-momentum4}
\end{subequations}
And for the energy:
\begin{equation}
\boxed{
\left[ \rho_g  (u h - u_g e)   - (\tau_{xx} u + \tau_{xy} v +\tau_{xz} w) + \dot{q}''_{x} + \sum J_{x,i} h_i \right]_f = \left[ - \rho_s u_g e_s +  \dot{q}''_{x}   \right]_s
}
\end{equation}

It should be noted that these equations are related to boundary conditions often found in the relevant literature \cite{Mortensen2016,Trevino2015aa}. For instance, by neglecting the gas velocity at the wall, these equations reduce to:
\begin{equation}
{
\left[ - \rho_{g_i} u_g +J_{x,i}  \right]_f = \left[ -\rho_{i} u_g  \right]_s
}
\label{consofmass4final}
\end{equation} 
\begin{subequations}
\begin{align}
\left[  p  -\tau_{xx} \right]_f  &=  R''_x\\
\left[   -\tau_{yx} \right]_f  &= R''_y\\
\left[   -\tau_{zx} \right]_f  &= R''_z
\end{align}
\label{eqn:full-momentum4final}
\end{subequations}
\begin{equation}
{
\left[ -\rho_g   u_g e   + \dot{q}''_{x} + \sum J_{x,i} h_i \right]_f = \left[ - \rho_s u_g e_s +  \dot{q}''_{x}   \right]_s
}
\end{equation}

\section{Conclusion} %
It is demonstrated that complex boundary conditions need not be derived one example at a time, a process 
that is normally ``painful" and time consuming. 
Rather, a systematic and universal method of matching governing equations at the boundaries and interfaces can provide a rigorous approach that can consistently result in the correct set of boundary and interface conditions. This method is successfully tested against four well-known sets of flow conditions, namely Beavers-Joseph flow, an aerothermodynamic interface flow, an adiabatic impermeable wall, and an ablating surface. 

It is expected that when applied to other flow configurations, independent of the complexity of the flows, the underlying method presented here will  result in an accurate and useful set of boundary and interface conditions. 
\section{Acknowledgments}
Financial support for this work was provided by NASA Award NNX15AD73G and NNX13AN04A.

\bibliography{pain}
\end{document}